\documentclass[%
 reprint,
latex apssamp.tex,
bibtex apssamp
bibnotes,
 amsmath,amssymb,
 aps,
pra,
]{revtex4-2}

\usepackage{xcolor}
\usepackage[colorlinks=true,pdfborder={0 0 0},linkcolor=blue]{hyperref}
\usepackage{caption}
\usepackage{subcaption}
\usepackage{graphicx}
\usepackage{dcolumn}
\usepackage{bm}
\usepackage{tikz}

\begin{document}

\preprint{APS/123-QED}

\title{$\mathbf{n}$-qubit states with maximum entanglement across all $\mathbf{k}$ vs $\mathbf{n-k}$ bi-partitions: A graph state approach}

\author{Sowrabh Sudevan}
\email{ss18ip003@iiserkol.ac.in}
\author{Sourin Das}%
\email{sourin@iiserkol.ac.in}
\affiliation{ Indian Institute of Science Education and Research Kolkata
\\Mohanpur, Nadia-741 246, West Bengal, India}

\date{\today}

\begin{abstract}
We discuss the construction of $n$-qubit pure states with maximum bipartite entanglement across all possible choices of $k$ vs $n-k$ bi-partitioning, which implies that the Von Neumann entropy of every $k$-qubit reduced density matrix corresponding to this state should be $k \ln 2 $. Such states have been referred to as $k$-uniform, $k$-MM states. We show that a subset of the '{\it{graph states}}' satisfy this condition, hence providing a recipe for constructing $k$-uniform states. Finding recipes for construction of $k$-uniform states using graph states is useful since every graph state can be constructed starting from a product state using only controlled-$Z$ gates.
Though, a priori it is not clear how to construct a graph which corresponds to an arbitrary $k$-uniform state, but in particular, we show that graphs with no isolated vertices are $1$-uniform. Graphs organized as a circular linear chain corresponds to the case of $2$-uniform state, where we show that the minimum number of qubits required to host such a state is $n=5$. $3$-uniform states can be constructed by forming bi-layer graphs with $n/2$ qubits ($n=2\mathbb{Z}$) in each layer, such that each layer forms a fully connected graph while inter-layer connections are such that the vertices in one layer has a one to one connectivity to the other layer. $4$-uniform states can be formed by taking 2D lattice graphs( also referred elsewhere as a 2D cluster Ising state ) with periodic boundary conditions along both dimensions and both dimensions having at least $5$ vertices.

\end{abstract}

\maketitle

\section{\label{sec:level1}Introduction:}

The physical phenomena of entanglement was discussed by Einstein, Podolsky, and Rosen in the famous paper Ref.~\cite{PhysRev.47.777} in 1935. Since then, there has been extensive discussions of various measures of entanglement \cite{Horodecki_2009,nielsen2002quantum,plenio2014introduction}. The entanglement measure this paper uses is bipartitite entanglement. We are interested in constructing qubit states that maximise the bipartite entanglement across all bipartitions of given number of qubits\cite{2004}. 
Page \cite{PhysRevLett.71.1291} showed that the most typical large qubit state has very high entanglement across the largest bipartition, i.e., it is trivial to see high bipartite entanglement for large $n$ qubit systems. Our work focusses instead on a finite number of qubits. We constructed recipes for $1,2,3$ and $4$-uniform qubit states across many different system sizes. Being able to engineer specific amounts of entanglement into a smaller number of qubits could be useful in quantum computation\cite{PhysRevLett.77.2585} and hence our findings could be relevant in this context.\\ 
$k$-uniform states have been studied in the past \cite{Facchi_2008,Arnaud_2013,2018}. 
Arnaud et al.\cite{Arnaud_2013} expanded their pure state density matrices in the generalized Bloch basis and connected $k$-uniformity to the presence and absence of certain types of Bloch expansions terms and then connected this to the quantum Gilbert-Varshamov and quantum Hamming
bounds \cite{PhysRevLett.77.2585} to claim a lower bound on the amount of entanglement that can be seen in a typical state with large number of qubits. Facchi et al.\cite{Facchi_2008} worked in the qubit basis to prove the existence of an extremal version of $k$-uniformity (absolutely maximally entangled states\cite{PhysRevA.86.052335,2014,PhysRevLett.128.080507}) for small number of qubits and also disprove the existence of such absolutely maximally entangled states in n qubits systems with $n\geq 8$. Most of the states we have constructed here do not approach this extremal case. \\
One of the motivations for our work is to identify recipes i.e., quantum circuits, that when applied to a product state leads to $k$-uniform states. $n$ qubit graph states in their pictorial representation come with the quantum circuit for constructing them starting from the product state $|+\rangle^{\otimes n}$ \cite{hein2006entanglement} and hence it is natural to search within graph states for $k$-uniformity, which is the main focus of this work.\\
 There exists a one to one map from any simple graph\cite{west2001introduction}( undirected, unweighted and without multiple edges or loops) to a pure $n$ qubit state. The graph edges also tell us what unitary operations need to be applied to realise these states. Hence finding a recipes for construction of $k$-uniform states using graph state automatically paves the way to identification of unitary transformations which when apply to a product state will generate these $k$-uniform states. Graph states have been experimentally realised \cite{Lee:12} and it has been shown that these states exhibit rich entanglement structures \cite{Shenoy_2019,Hein_2004,Gisin_1998,Adcock2020mappinggraphstate}, have applications in quantum computation for a different model of computation using measurements on graphs states\cite{PhysRevLett.86.5188} and their local unitary equivalents are also useful in quantum error correction\cite{gottesman1997stabilizer}.

\section{\label{sec:level1}Maximally entangled states:}
An $n$ qubit state is $k$-uniform, if any $k$ qubit reduced density matrix of this state is maximally mixed.
It is straightforward to show that a $k$-uniform state is also a $1,2,3,\dots,(k-1)$-uniform state\cite{nielsen2002quantum}. 
An $n$ qubit state can at most be $\left\lfloor \frac{n}{2}\right\rfloor $-uniform, which follows from constraints on the dimensions of the reduced density matrix and the fact that both reduced density matrices in the bi-partition should have the same entropy. An $n$ qubit state is called absolutely maximally entangled \cite{helwig2013absolutely} if it is $\left\lfloor \frac{n}{2}\right\rfloor $-uniform.

For example, a class of $1$-uniform states is given by\cite{Arnaud_2013}, \begin{eqnarray}
\frac{1}{\sqrt{(2)}}\left(|0\rangle^{\otimes n}+|1\rangle^{\otimes n}\right)
\end{eqnarray}. For $n=2$, this corresponds to the well known Bell state\cite{sakurai1995modern} and for $n\geq3$ this corresponds to generalized GHZ states\cite{nielsen2002quantum,greenberger2007going}.

The $3$ qubit GHZ state in particular turns out to be absolutely maximally entangled, since $1=\left\lfloor \frac{3}{2}\right\rfloor$. Generalized GHZ states with $n>3$ are not absolutely maximally entangled. 
These k-uniform states can also be constructed by orthogonal arrays[\cite{PhysRevA.90.022316,2019npjQI...5...52P,2021JPhA...54a5305P,PhysRevA.99.042332,7918542}]
\subsection{\label{sec:level2}Generalized Bloch basis}
In this section we introduce the generalized Bloch basis  \cite{Arnaud_2013,PhysRev.70.460,nielsen2002quantum} for describing an $n$ qubit system.
In this paper we have restricted our study to systems comprising of $n$ qubits, hence our density matrices($\rho$) are of dimensions  $2^{n} \times 2^{n}$. Since $\rho$ is also Hermitian, it can be expanded in terms of tensor products of Pauli matrices($\sigma_{0}= I =\left[\begin{array}{cc}
1 & 0\\
0 & 1
\end{array}\right],\sigma_{1}= X = \left[\begin{array}{cc}
0 & 1\\
1 & 0
\end{array}\right],\sigma_{2}= Y = \left[\begin{array}{cc}
0 & -i\\
i & 0
\end{array}\right],\sigma_{3}= Z = \left[\begin{array}{cc}
1 & 0\\
0 & -1
\end{array}\right]$),
\begin{eqnarray}
\rho_{{2^{n}}\times {2^{n}}}=\sum_{i_{1},i_{2}...,i_{n}}c_{i_{1},i_{2},...,i_{n}}\sigma_{i_{1}}\otimes\sigma_{i_{2}}\otimes...\otimes\sigma_{i_{n}}
\end{eqnarray}
where,
\begin{align}
c_{i_{1},i_{2},i_{3},...,i_{n}}=\frac{1}{2^{n}}Trace(\sigma_{i_{1}}\otimes\sigma_{i_{2}}\otimes\sigma_{i_{3}}\otimes...\otimes\sigma_{i_{n}}\times\rho)
\end{align}
For example, in this basis the Bell state, $|\psi\rangle=\frac{1}{\sqrt{2}}(|01\rangle+|10\rangle)$ can be expanded as,
\begin{eqnarray}
\left(|\psi\rangle\right)(|\langle\psi|)=\frac{1}{4}(I\otimes I+X\otimes X+Y\otimes Y-Z\otimes Z)
\end{eqnarray}
For the rest of the paper we will omit the explicitly written tensor product symbols and then, the typical terms appearing in an expansion are 'Pauli words' like, $XX, YY, ZZXIIIX$ and so on. These Pauli words can be classified as $1$-particle, $2$-particle $\dots$ n-particle terms , based on how many non identity matrices they contain. Example: $XXY$ is a $3$-particle term, $XIZZIY$ is a $4$-particle term and so on. Now we define the weight of a Pauli word, which is equal to the number of non identity Pauli matrices it contains.
At this point one should remember $c_{i_{1},i_{2},i_{3},\dots,i_{n}}$ also corresponds to the expectation value of an operator $\sigma_{i_{1}}\sigma_{i_{2}}\sigma_{i_{3}}\dots\sigma_{i_{n}}$  in state $\rho$ is also given by the same expression. Therefore expanding density matrices in this basis, directly gives us access to the average values of various physical observables associated with the individual spins for systems in that state.
We start by making the following claim:\\
\\
\textit{The expectation value of all one particle operators is zero for $1- $uniform states \footnote{Though this claim was stated and proven in \cite{Arnaud_2013}, we provide an alternate proof here for completeness.}}\\
\\
\textit{Proof:}
 From (3), we compute $\displaystyle c_{0,0,\dots,0}=\frac{1}{2^{n}}Trace(I_{2^{n}\times2^{n}}\times\rho)=\frac{1}{2^{n}}$. Note that the rest of the terms is a subset of the set of $1$-particle terms, $2$-particle terms$\dots$ $n$-particle terms.
 Therefore the first term of the Bloch expansion of any state is $\frac{1}{2^{n}} I_{2^{n} \times 2^{n}}$. Partial tracing is distributive across matrix addition, so we can sum over the partial traces of the different matrices in the Bloch expansion.
 Partial tracing out $(n-1)$ qubits from $\frac{1}{2^{n}} I_{2^{n} \times 2^{n}}$ can give us our required fully mixed $\frac{1}{2} I_{2 \times 2}$ matrix. Therefore, we need the rest of the matrices in the Bloch expansion of $\rho$ to partial trace out to zero.\\
 Partial tracing distributes across matrix tensor products as matrix multiplications. example: $Trace_{23}(A_{1}\otimes B_{2}\otimes C_{3})=A_{1}\times Trace(B_{2})\times Trace(C_{3})$
 An $m$-particle term in the Bloch expansion of an $n$-qubit state has $m$ zero-trace matrices($X,Y,Z$) and $(n-m)$ Identities($I$). For $m\geq2$, when we partial trace out $(n-1)$ qubits, we will be forced to include at least one $trace=0$ matrix in the set of matrices being traced over by $\textit{pigeon hole principle}$. This will make that term in the Bloch expansion disappear and so we are left with only the $m=1$ and $m=0$ operators. These correspond to the $1$-particle and $\frac{I_{2^{n}\times2^{n}}}{2^{n}}$. So we can see that if a Bloch expansion does not contain any $1$-particle terms then, all terms other than $\frac{I_{2^{n}\times2^{n}}}{2^{n}}$ will partial trace out to zero and hence we will be left with the fully mixed $\frac{I_{2\times2}}{2}$ matrix.
 This implies $1$-uniform states do not contain one particle terms in their Bloch expansions. This idea of the lack of one particle operators can be extended in a straightforward way to make the following more useful result.\\
\\
 \underline{\textit{Theorem 1}:} The expectation values/coefficients in the Bloch expansions of $1,2,3,\dots, k$-particle terms will be zero for  $k$-uniform states.\\
 \\
For an n qubit pure state density matrix to become a k qubit maximally mixed density matrix upon tracing out n-k qubits, we need the absence of any 1,2,3,..k-particle operators in its Bloch expansion. 1,2,3,..k-particle operators are the only operators that may not trace out to zero, when tracing out n-k qubits from an n qubit state. All higher particle operators will necessarily partially trace out to zero.

\subsection{\label{sec:level2}Graph states}

We start this section by reviewing the definition of graph states for the sake of completeness.
Graphs \cite{west2001introduction} are a collection of vertices and edges(the vertices are connected by the edges), which can be represented as $G=(V,E)$ where, $V$ is the set of vertices and $E$ is the set of edges respectively. In the graphs that we consider, there will be no self loops(an edge from a vertex to itself) and between any two vertices there can be at most one edge. Also we only consider undirected graphs (graphs where the edges are not assigned a sense of direction).
\begin{figure}
    \centering
    
    \label{fig:Example of 4 qubit graph}
    \begin{tikzpicture}

\draw[fill=black] (0,0) circle (3pt);
\draw[fill=black] (3,1) circle (3pt);
\draw[fill=black] (4,2) circle (3pt);
\draw[fill=black] (2,2) circle (3pt);
\node at (-1,0) {1,
XZII};
\node at (2,1) {2,ZXZZ};
\node at (5,2) {3,IZXI};
\node at (1,2) {4,IZIX};
\draw[thick] (0,0) -- (3,1) -- (4,2); 
\draw[thick] (2,2) -- (3,1);
\end{tikzpicture}
\caption{Example of 4 qubit graph}
\label{4_qubit_graph}    
\end{figure}
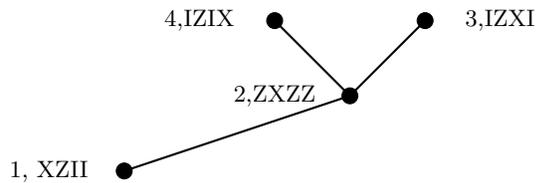

Corresponding to any undirected n vertex graph as in figure \ref{4_qubit_graph}, we can write down an n qubit pure quantum state called a graph state \cite{hein2006entanglement}. 
The vertices are labelled with certain operators (called correlation operators) and it should be noted that these operators commute. The way these correlation operators are constructed is by first numbering each vertex as representing a qubit. There is one correlation operator per graph vertex. The qubit states we are looking at are not identical in the quantum mechanical sense and so in an n qubit system, each qubit can be numbered. Once we number the vertices/qubits of the graph the correlation operator corresponding to each vertex can be constructed by including a Pauli X($\sigma_{x}$) acting on the Hilbert space of the i'th qubit and Pauli Z($\sigma_{z}$) matrices at the Hilbert spaces of the qubits that are connected to the i'th qubit, and Pauli I(Identity) matrices elsewhere. For example, in the graph in figure \ref{4_qubit_graph}, the correlation operators are given by $XIZZ$, $IXZI$, $ZZXZ$, $ZIZX$. These four operators commute and square to identity and these two properties allow the correlation operators to be the generators of a $16$ element ${Z_{2}}^4$ abelian group of operators under matrix multiplication. In general the n correlation operators of an n vertex graph generates $2^{n}$ operators called the stabilizer of the state. The sum of all  these stabilizer operators multiplied by $\frac{1}{2^{n}}$ gives the density matrix for the n qubit graph state. The stabilizers of an n qubit state form a ${Z_{2}}^n$ group.
 
As an example we can write the full expansion of the graph in fig \ref{4_qubit_graph}, $|G_{4}\rangle$,
\begin{multline}
 |G_{4}\rangle\langle G_{4}|=\frac{1}{16}(IIII+XZII+ZXZZ+IZXI+IZIX\\+YYZZ+ZYYZ+IIXX+XIXI+ XIIX+ZYZY\\-YXYZ-ZXYY+XZXX-YXZY-YYYY)   
\end{multline}

\section{\label{sec:level3}\textbf{Entanglement in graph states}:}
There have been previous works that have explored the entanglement content of graph states  through the use of different entanglement monotones \cite{Hein_2004,Hajdu_ek_2013}. There has also been previous work exploring entanglement in the same spirit as this paper, where they find a lower bound on the value of k for k-uniform states that exist as large graph states using connections to quantum error correction as well as establishing that permutation symmetric stabilizer states \cite{Arnaud_2013}  can be at most 1-uniform. In contrast we are considering graph states that are not permutation symmetric as a possible construction of k-uniform states. To make progress we start with the following definition.\\ 
\\
  \underline{\textit{Definition:}} k-products is the set of elements generated by products between k correlation operators of a graph state.
  The different k-product sets are disjoint subsets of the full Bloch expansion because the correlation operators are elements of an abelian group. Now note that the smallest particle sector in the Bloch expansion of a k-uniform n qubit state is $k+1$. The weight of a correlation operator corresponding to a vertex with degree $d$ is $d+1$. The correlation operators are themselves part of the Bloch expansion and hence a k-uniform graph state will only contain vertices with $degree,d\geq k$.
  
  While $degree,d \geq k$ for every vertex of a graph is necessary for the state to be k-uniform, this is not sufficient because it is possible for products of correlation operators to have smaller weight. Since we need at least degree k for each vertex a useful starting point in a search for k-uniform graphs could be to look at k-regular graphs(graphs with $degree=k$ for each vertex).
 
 (Since we are interested in studying the weights of different Bloch expansion terms)
Note that the i'th correlation operator can be written as,
\begin{eqnarray}
K_{i}=X_{i}\otimes\displaystyle\prod_{\otimes l}^{l\in N_{i}}Z_{l}
\end{eqnarray}
 where $i\in {1,2,3,...n}$ for n qubit system.$N_{i}$, the neighbourhood of vertex i, is the set of vertices that are connected to the vertex i with exactly one edge. The set of $N_{i}$'s comes from the underlying graph. For example, in our previous 4 qubit graph(figure 1), $N_{4} = [1,3]$.  
and hence $K_{4}=X_{4}Z_{1}Z_{3}$.

To understand what a product of k correlation operators looks like,let us first construct a general two-product between two correlation operators $K_{i}$ and $K_{j}$ with $N_{i}$ $=$ neighbourhood of i and $N_{j}$ $=$ neighbourhood of j,

\begin{multline}
    K_{i}\times K_{j}=\left(X_{i}{\displaystyle \prod_{\otimes l}^{l\in N_i}Z_{l}}\right)\times\left(X_{j}{\displaystyle \prod_{\otimes m}^{m\in N_j}Z_{m}}\right)
\end{multline}
where,
\begin{multline}
   i\neq j\in N=\{1,2,3,..n\};l\subset N-\{i\};m\subset N-\{j\}
\end{multline}

which can be reduced to,
\begin{eqnarray}
K_{i}\times K_{j}=X_{i}X_{j}{\displaystyle \prod_{\otimes l}^{l\in S_{2}}Z_{l}}
\end{eqnarray}
where, $S_{2}=N_{i} \oplus N_{j}$. "$\oplus$" is the symmetric difference binary operation between sets, which is defined as, $A\oplus B=(A\cup B)-(A\cap B)$. Note that all $Z$'s will have distinct indices owing to the fact that $Z$'s belonging to the same index multiply with each other in pairs to give identity as $Z^{2} = I$. As defined earlier, the weight of the above product is equal to the number of distinct indices in the string of Pauli operators. We have unique indices in eqn 9 for the $Z$ operators. If we include $X_{i}$ and $X_{j}$ we do not have all unique indices if $S_{2}\cap\{i,j\}\neq\phi$, where $\phi$ denotes the empty set. Note that while $N_{i}$ cannot contain i and $N_{j}$ cannot contain j,  $N_{i}\oplus N_{j}$ can contain both.
To ensure that every index in eqn 9 is unique, we take out $Z_{i}$ and $Z_{j}$ if present, from the product over $Z$'s and subtract $\lbrace i,j\rbrace$ from the set $S_{2}$. These $Z$'s that we have taken out will multiply with $X_{i}$ and $X_{j}$ depending on whether they exist in the symmetric sum $S_{2}$ or not. Finally the two-product looks like,
\begin{eqnarray}
K_{i}\times K_{j}=X_{i}(Z_{i})^{C_{i}(S_{2})}X_{j}(Z_{j})^{C_{j}(S_{2})}{\displaystyle \prod_{\otimes l}^{l\in S_{2}-\{i,j\}}Z_{l}}
\end{eqnarray}
Where $C_{i}(S_{2})$ is the cardinality of i in $S_{2}$, which means the number of times element i appears in set $S_{2}$.
So here we have a neat way of writing a general 2-product where each Pauli operator has a unique index. We see that this is easily generalizable to 3-products and greater. A general 3-product would look like,
\begin{multline}
    K_{i}K_{j}K_{r}=X_{i}(Z_{i})^{C_{i}(S_{3})}X_{j}(Z_{j})^{C_{j}(S_{3})}\\{\displaystyle X_{r}(Z_{r})^{C_{r}(S_{3})}\prod_{\otimes l}^{l\in S_{3}-\{i,j,r\}}Z_{l}}
\end{multline}
where $S_{3}=N_{i}\oplus N_{j}\oplus N_{r}$

The general n-product would look like,
\begin{multline}
K_{i_{1}}K_{i_{2}}\dots K_{i_{n}}=X_{i_{1}}(Z_{i_{1}})^{C_{i_{1}}(S_{n})}X_{i_{2}}(Z_{i_{2}})^{C_{i_{2}}(S_{n})}\dots\\X_{i_{n}}(Z_{i_{n}})^{C_{i_{n}}(S_{n})}{\displaystyle \prod_{\otimes l}^{l\in S_{n}-\{i_{1},i_{2},...,i_{n}\}}Z_{l}}
\end{multline}
where $S_{n}=N_{i_{1}}\oplus N_{i_{2}}\oplus N_{i_{3}}\oplus\dots\oplus N_{i_{n}}$,

Looking at (10), (11) and (12), we see that the weight of these products regardless of the weight of what appears inside the product($\displaystyle\prod$) is at least equal to the number of $X$'s in front of the $\displaystyle\prod$. The number of $X$'s is clearly equal to the number of different correlation operators being multiplied, therefore in general,
\\
\\
\underline{\textit{Theorem 2}:} $w(k-product)\geq k$.  k-product has minimum weight of k.
\\

The theorem above is very important to the rest of this paper. What we have shown is that a 2-product has minimum weight of two and a 3-product has a minimum weight of 3 and so on, regardless of what graph you take. This result allows us to avoid computing the full Bloch expansion that corresponds to a given graph. We only need to compute a certain number of terms depending on what entanglement we expect in the state. With the n correlation operators we can read off of any given graph, one has to compute $2^{n}$ products to get the full Bloch expansion. But if one only needs to check if the state is k-uniform, since $k\leq\left\lfloor \frac{n}{2}\right\rfloor $ we only need to evaluate $\left(\begin{array}{c}
n\\
0
\end{array}\right)+\left(\begin{array}{c}
n\\
1
\end{array}\right)+\left(\begin{array}{c}
n\\
2
\end{array}\right)+\cdots+\left(\begin{array}{c}
n\\
k
\end{array}\right)\leq2^{n-1}$ terms.
So, for a state to be 1-uniform, we only need weights of 1-products to be more than one, for a state to be 2-uniform we only need weights of 1-products and 2-products to be more than two, for a state to be 3-uniform we only need weights of 1-products, 2-products and 3-products to be more than three. We do not need to evaluate any of the other terms of the Bloch expansion when checking for 1,2 or 3$\dots$-uniformity in graph states.
 
 \subsection{\label{sec:level3}1 - uniform states:}
\begin{figure}
     \centering
     \begin{subfigure}[b]{0.3\textwidth}
             \centering

    \begin{tikzpicture}

\draw[fill=black] (1,0) circle (3pt);
\draw[fill=black] (0,1) circle (3pt);
\draw[fill=black] (1,1) circle (3pt);
\draw[fill=black] (2,1) circle (3pt);
\draw[fill=black] (0,2) circle (3pt);
\draw[fill=black] (1,2) circle (3pt);
\draw[fill=black] (2,2) circle (3pt);

\node at (0.5,0) {1};
\node at (-0.5,1) {2};
\node at (1,1.5) {3};
\node at (-0.5,2) {4};
\node at (1,2.5) {5};
\node at (2.5,2) {6};
\node at (2.5,1) {7};

\draw[thick] (1,0) -- (1,1) -- (0,1);
\draw[thick] (0,2) -- (1,2);
\draw[thick] (2,2)--(2,1);
\end{tikzpicture}
\caption{7 qubit 1-uniform state}
\label{7 qubit 1-uniform state}
\end{subfigure}
     
\begin{subfigure}[b]{0.3\textwidth}
\centering

\begin{tikzpicture}

\draw[fill=black] (0,0) circle (3pt);
\draw[fill=black] (3,0) circle (3pt);
\draw[fill=black] (3,3) circle (3pt);
\draw[fill=black] (0,3) circle (3pt);
\node at (-0.5,0) {1};
\node at (3.5,0) {2};
\node at (3.5,3) {3};
\node at (-0.5,3) {4};
\draw[thick] (0,0) -- (3,0) -- (3,3) -- (0,3) -- (0,0)--(3,3);
\draw[thick] (0,3)--(3,0);
\end{tikzpicture}
\caption{4 qubit fully connected graph state}
\label{fig:4 qubit fully connected graph state}
\end{subfigure}
     \hfill
\caption{1-uniform states}
\label{1-uniform states}
\end{figure}
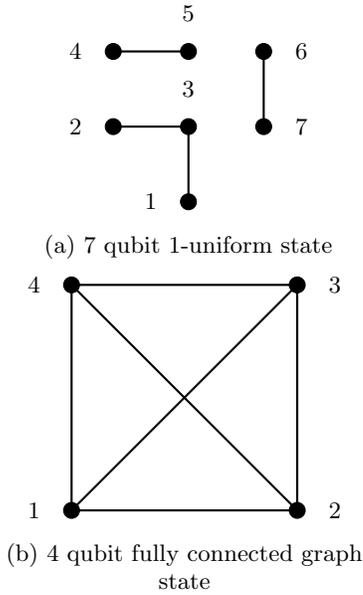

 As per theorem 1 we need the correlation operators to have a weight of at least 2 $(w\geq2)$ in a 1-uniform state. Because of theorem 2 we only need to ensure the 1-products(correlation operators) of 1-uniform states have weight more than one. This can be easily ensured by connecting every vertex to at least one other vertex.
So the recipe for a 1-uniform state is :\\ 
\\
  \textit{Graphs where the degree of every vertex is greater than zero are 1-uniform.}\\ 
  \\
 If one needs to construct a 1-uniform state with a minimum number of controlled-Z unitaries applied to a product state, then one needs to pair off all the vertices(qubits). For an odd number of qubits, we need one extra edge to connect the left out odd qubit with one of the pairs(figure \ref{7 qubit 1-uniform state}).

\textit{n-qubit GHZ state is 1-uniform}:
N qubit GHZ state is equivalent to an n qubit fully connected graph state up to local unitary transformations \cite{hein2006entanglement}. A fully connected graph is one, where every vertex is connected with an edge each, to every other vertex as in figure \ref{fig:4 qubit fully connected graph state}

Now comparing this graph with the result in the previous section, we know the GHZ state is at least 1-uniform. Now we show how it is exactly 1-uniform.

Each of the correlation operators have weight n for n qubit complete graph. Now look at the two products,
\begin{eqnarray}
K_{i} K_{j}=X_{i}(Z_{i})^{C_{i}(S_{2})}X_{j}(Z_{j})^{C_{j}(S_{2})}{\displaystyle \prod_{\otimes l}^{l\in S_{2}-\{i,j\}}Z_{l}}
\end{eqnarray}
All vertices are connected to all other vertices,
\begin{eqnarray}
N_{i}\cup\{i\}=N_{j}\cup\{j\}
\end{eqnarray}
\begin{eqnarray}
\implies S_{2}-\{i,j\}=N_{i}\oplus N_{j}-\{i,j\}=\phi
\end{eqnarray}
$\phi$ is the null set. So the weight of what's inside the product in eqn(13) is zero. Therefore all two products of GHZ states are two particle terms. By theorem 2 Three and higher products have higher weights, therefore the smallest term in the Bloch expansion of the state has weight two and hence an n qubit GHZ is exactly 1-uniform.

\subsection{\label{sec:level3}2 - uniform states:}
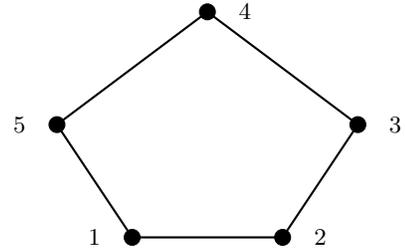
\begin{figure}
    \centering

    \begin{tikzpicture}

\draw[fill=black] (1,0) circle (3pt);
\draw[fill=black] (3,0) circle (3pt);
\draw[fill=black] (4,1.5) circle (3pt);
\draw[fill=black] (2,3) circle (3pt);
\draw[fill=black] (0,1.5) circle (3pt);
\node at (0.5,0) {1};
\node at (3.5,0) {2};
\node at (4.5,1.5) {3};
\node at (2.5,3) {4};
\node at (-0.5,1.5) {5};
\draw[thick] (1,0) -- (3,0) -- (4,1.5) -- (2,3) -- (0,1.5)--(1,0);

\end{tikzpicture}
\caption{5 qubit 2-uniform state}
\label{5 qubit 2-uniform state}
    
\end{figure}
To construct 2-uniform states we need to follow the below recipe.\\
 \\
 \textit{ Qubit state with more than four qubits arranged in a circular linear chain is $2$-uniform}\\
\\
 We can start with $4$-qubit systems, since smaller systems cannot have $2$-uniformity (due to the symmetry of the entropy in the two parts after bi-partitioning  as well as the fact that the maximum possible entropy of a $2^{n}\times2^{n}$ matrix is $n\log2$). We take the corresponding correlation operators for a "closed chain graph" as our ansatz and show that it is indeed $2$-uniform. The general form of the correlation operators for a closed chain graph is given by  
\begin{eqnarray}
K_{i}=Z_{i-1}X_{i}Z_{i+1}.
\end{eqnarray}
For five qubits this ansatz gives us figure \ref{5 qubit 2-uniform state}.

Now these are three particle terms. The $i+1$ and $i-1$ are sums modulo $n$ with qubits being numbered along the chain starting from $0$ upto $(n-1)$. Example, for 4 qubits, $(3+1)mod(4) = 0$ and $(1-1)mod4 = 0$.

Two-product looks like ,
\begin{eqnarray}
K_{i}K_{j}=Z_{i-1}X_{i}Z_{i+1}Z_{j-1}X_{j}Z_{j+1}
\end{eqnarray}
If $i + 1 = j$, that is they are neighbours,
\begin{eqnarray}
K_{i}K_{i+1}=Z_{i-1}Y_{i}Y_{i+1}Z_{i+2}
\end{eqnarray}
This is a four body term

if $i+2 = j$, that is they are next nearest neighbours,
\begin{eqnarray}
K_{i}K_{i+1}=Z_{i-1}X_{i}X_{i+2}Z_{i+3}
\end{eqnarray}
This is also a four body term.
For $4$ qubits, $(i-1)mod(4)=(i+3)mod(4)$ and hence $Z_{i-1}$ will square to identity with $Z_{i+3}$ and therefore, for $4$ qubits, $2$ product between next nearest neighbours becomes two particle term and hence $4$ qubit state following this recipe cannot be $2$-uniform. But for any qubit state above four qubits, this cancellation due overlapping of indices from modular addition does not take place and we again get four particle terms.
All two products other than those between nearest neighbours and next nearest neighbours are between correlation operators with no overlap, and hence they give $6$ particle terms,
\begin{eqnarray}
K_{i}K_{j}=Z_{i-1}X_{i}Z_{i+1}Z_{j-1}X_{j}Z_{j+1}
\end{eqnarray}
 \begin{figure*}
     \centering
     \begin{subfigure}[b]{0.3\textwidth}
         \centering
         \begin{tikzpicture}[scale = 2.25]

\draw[fill=black] (0,0) circle (1.5pt);
\draw[fill=black] (1.5,0) circle (1.5pt);
\draw[fill=black] (0.75,1) circle (1.5pt);
\draw[fill=black] (0,1.5) circle (1.5pt);
\draw[fill=black] (1.5,1.5) circle (1.5pt);
\draw[fill=black] (0.75,2.25) circle (1.5pt);
\node at (-0.25,0) {1};
\node at (1.75,0) {2};
\node at (1,1) {3};
\node at (-0.25,1.5) {4};
\node at (1.75,1.5) {5};
\node at (1,2.25) {6};
\draw[thick] (0,0) -- (1.5,0) -- (0.75,1) -- (0,0) -- (0,1.5)--(1.5,1.5)--(0.75,2.25)--(0,1.5);
\draw[thick](0.75,2.25)--(0.75,1);
\draw[thick](1.5,1.5)--(1.5,0);

\end{tikzpicture}
         \caption{$\left[\begin{array}{cccccc}
X & Z & Z & Z & I & I\\
Z & X & Z & I & Z & I\\
Z & Z & X & I & I & Z\\
Z & I & I & X & Z & Z\\
I & Z & I & Z & X & Z\\
I & I & Z & Z & Z & X\\
\end{array}\right]$}
         \label{fig:$6$ qubit $3$-uniform state}
     \end{subfigure}
     \hfill
     \begin{subfigure}[b]{0.3\textwidth}
           \centering

    \begin{tikzpicture}[scale = 2]

\draw[fill=black] (0,0) circle (1.5pt);
\draw[fill=black] (1.5,0) circle (1.5pt);
\draw[fill=black] (1.75,0.75) circle (1.5pt);
\draw[fill=black] (0.25,0.75) circle (1.5pt);
\draw[fill=black] (0,1.5) circle (1.5pt);
\draw[fill=black] (1.5,1.5) circle (1.5pt);
\draw[fill=black] (1.75,2.25) circle (1.5pt);
\draw[fill=black] (0.25,2.25) circle (1.5pt);
\node at  (-0.25,0) {1};
\node at (1.75,0) {2};
\node at (2,0.75) {3};
\node at (0.125,0.75) {4};
\node at (-0.25,1.5) {5};
\node at (1.625,1.5) {6};
\node at (2,2.1) {7};
\node at (0,2.25) {8};
\draw[thick] (0,0) -- (1.5,0) -- (1.75,0.75) -- (0.25,0.75) -- (0,0)--(1.75,0.75);
\draw[thick] (0.25,0.75)--(1.5,0);
\draw[thick] (0,1.5)--(1.5,1.5)--(1.75,2.25)--(0.25,2.25)--(0,1.5);
\draw[thick] (0,1.5)--(1.75,2.25);
\draw[thick] (0.25,2.25)--(1.5,1.5);
\draw[thick] (0,1.5)--(0,0);
\draw[thick] (0.25,2.25)--(0.25,0.75);
\draw[thick] (1.75,0.75)--(1.75,2.25);
\draw[thick] (1.5,1.5)--(1.5,0);

\end{tikzpicture}
\caption{$\left[\begin{array}{cccccccc}
X & Z & Z & Z & Z & I & I & I\\
Z & X & Z & Z & I & Z & I & I\\
Z & Z & X & Z & I & I & Z & I\\
Z & Z & Z & X & I & I & I & Z\\
Z & I & I & I & X & Z & Z & Z\\
I & Z & I & I & Z & X & Z & Z\\
I & I & Z & I & Z & Z & X & Z\\
I & I & I & Z & Z & Z & Z & X
\end{array}\right]$}
\label{fig:$8$ qubit $3$-uniform state}
     \end{subfigure}
     \hfill
     \begin{subfigure}[b]{0.3\textwidth}
         \centering

    \begin{tikzpicture}[scale = 0.7]

\draw[fill=black] (1,0) circle (3pt);
\draw[fill=black] (3,0) circle (3pt);
\draw[fill=black] (4,1.5) circle (3pt);
\draw[fill=black] (2,2) circle (3pt);
\draw[fill=black] (0,1.5) circle (3pt);
\draw[fill=black] (1,3) circle (3pt);
\draw[fill=black] (3,3) circle (3pt);
\draw[fill=black] (4,4.5) circle (3pt);
\draw[fill=black] (2,5) circle (3pt);
\draw[fill=black] (0,4.5) circle (3pt);
\node at (0.5,0) {1};
\node at (3.5,0) {2};
\node at (4.5,1.5) {3};
\node at (2.5,2.2) {4};
\node at (-0.5,1.5) {5};
\node at (0.5,3) {6};
\node at (3.5,3) {7};
\node at (4.5,4.5) {8};
\node at (2.5,5.2) {9};
\node at (-0.5,4.5) {10};
\draw[thick] (1,0) -- (3,0) --  (4,1.5) -- (2,2) -- (0,1.5)--(1,0)--(2,2)--(3,0)--(0,1.5)--(4,1.5)--(1,0);
\draw[thick] (1,3)--(3,3)--(4,4.5)--(2,5)--(0,4.5)--(1,3)--(2,5)--(1,3)--(0,4.5)--(4,4.5)--(1,3);
\draw[thick] (0,4.5)--(3,3)--(2,5);
\draw[thick](1,0)--(1,3);
\draw[thick](3,0)--(3,3);
\draw[thick](4,1.5)--(4,4.5);
\draw[thick](2,2)--(2,5);
\draw[thick](0,1.5)--(0,4.5);

\end{tikzpicture}
\caption{$\left[\begin{array}{cccccccccc}
X & Z & Z & Z & Z & Z & I & I & I & I\\
Z & X & Z & Z & Z & I & Z & I & I & I\\
Z & Z & X & Z & Z & I & I & Z & I & I\\
Z & Z & Z & X & Z & I & I & I & Z & I\\
Z & Z & Z & Z & X & I & I & I & I & Z\\
Z & I & I & I & I & X & Z & Z & Z & Z\\
I & Z & I & I & I & Z & X & Z & Z & Z\\
I & I & Z & I & I & Z & Z & X & Z & Z\\
I & I & I & Z & I & Z & Z & Z & X & Z\\
I & I & I & I & Z & Z & Z & Z & Z & X
\end{array}\right]$}
\label{fig:$10$ qubit $3$-uniform state}
    
     \end{subfigure}
        \caption{$3$-uniform states with $6$,$8$ and $10$ qubits and their corresponding adjacency matrices. Note the two diagonal blocks in all the matrices correspond to the fully connected sub graphs/layers and the off diagonal blocks correspond to the one to one connections between the two layers.}
        \label{$3$-uniform states}
\end{figure*}
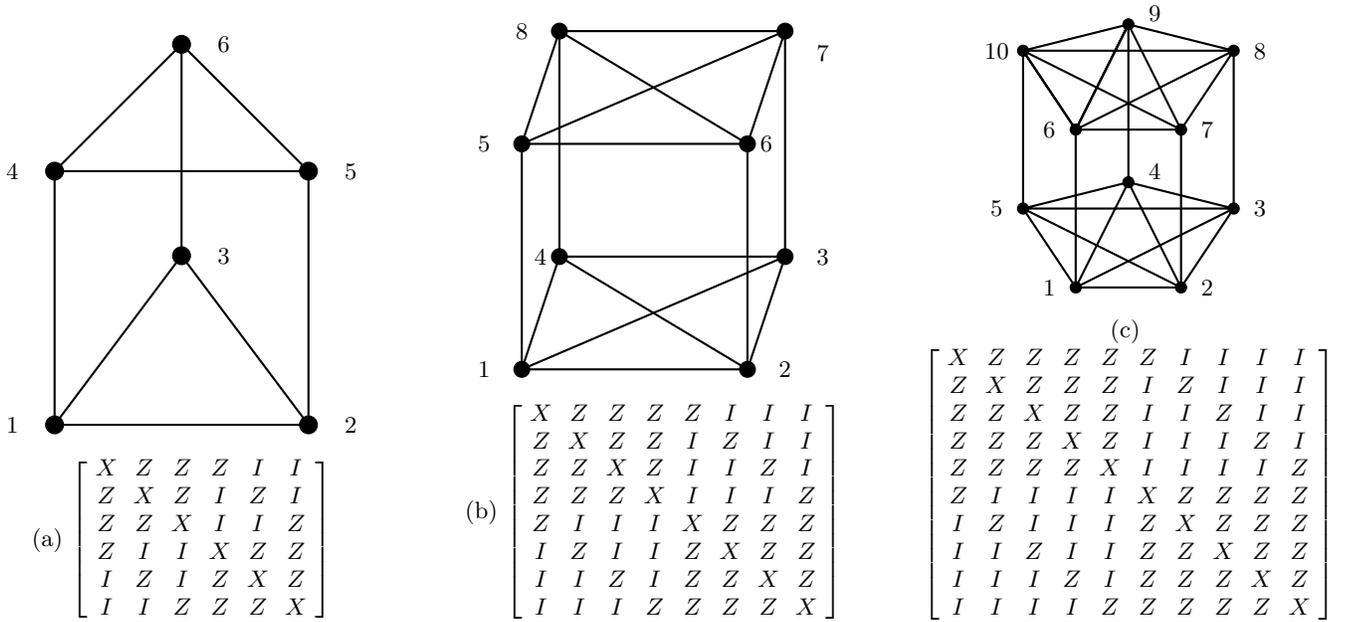
To summarise, two products are all giving at least 4 particle terms for systems with more than $4$ qubit in our ansatz, three products and higher give at least $3$ particle terms and higher due to theorem-$2$, and individual correlation operators are $3$-particle terms, therefore this recipe gives us $2$-uniform states.

 \subsection{\label{sec:level3}$3$ - uniform states:}

To construct a $3$-uniform state the recipe is given below,\\
\\
 \textit{$2n$ qubit state is split into a bi-layer graph, with $n$ qubits in each layer. Within a layer, all the qubits are fully connected to one another and between layers the edges form a one to one mapping between the two sets/layers of $n$ qubits.}\\
 \\
 The examples in figures \ref{fig:$6$ qubit $3$-uniform state} ,\ref{fig:$8$ qubit $3$-uniform state} ,\ref{fig:$10$ qubit $3$-uniform state} for $6$,$8$ and $10$ qubits might help clarify our construction.

This construction can be extended for any even number of qubits. To understand why this construction works we need to look at the adjacency matrices(fig:\ref{3-uniform states}) for these three graphs.
Adjacency matrices are ubiquitous in graph theory, for our purposes here, all we have done is arrange the correlation operators for our $3$-uniform graphs in rows, such that the Pauli $X$'s are all along the main diagonal. Note that this necessarily forms a symmetric matrix as we are considering undirected graphs only. A "$Z$" in row $i$ and column $j$ of the matrix denotes that there is an edge between qubit $i$ and qubit $j$. An "$I$" in row $i$ and column $j$ of the matrix denotes the absence of an edge between qubit $i$ and qubit $j$.

Observing the matrices in fig:\ref{$3$-uniform states} we can see that even as we scale up this system, there is a persistent structure to these adjacency matrices. There are two diagonal blocks which corresponds to each fully connected layer of the bi-layer and the two off diagonal blocks represent the one to one connections between the two layers. For $3$-uniform states we need to analyze $1$-products, $2$-products and $3$-products and ensure they have weight greater than three, all higher products are necessarily four body and higher as per theorem-$2$. 

Now we prove how our ansatz is 3-uniform. For any 2n qubit system constructed according to our recipe, 1-products all have weight $n+1$, where $n$ weight comes from connections in the plane and one from the connection between planes. Due to the symmetry of this system, there are exactly three qualitatively "different" 2-products and 3- products. 
\begin{figure*}[t]
    \centering
    \includegraphics[width = \textwidth]{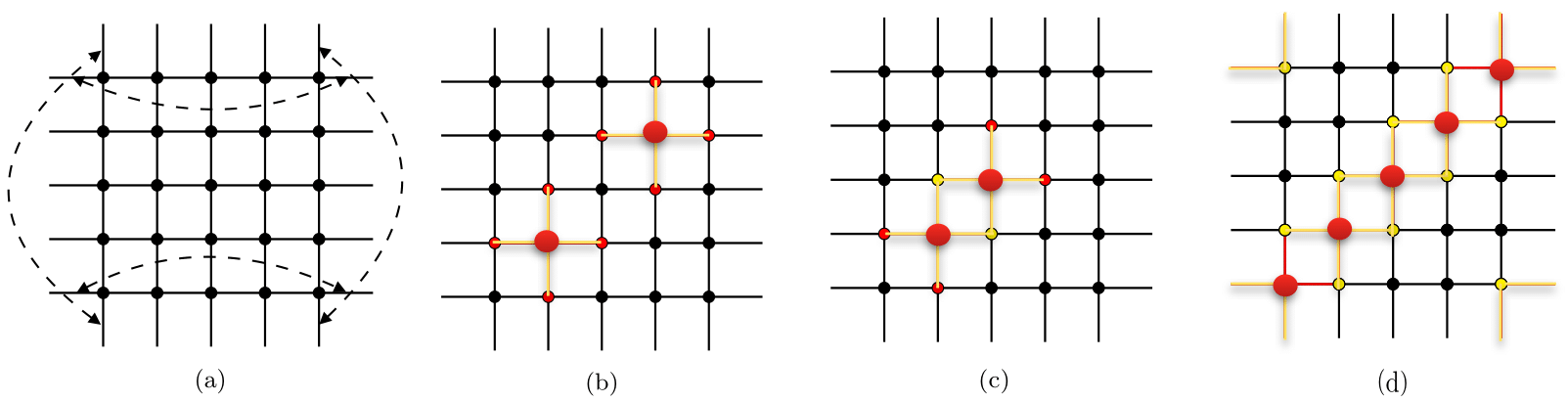}
    \caption{(a)25 qubit periodic lattice(torus) for 4-uniform state. Note that in this periodic lattice it is the edges outgoing from the diagramatically opposite ends along the two independent directions that are identified to form the torus. This is represented by the double headed arrows that represents the bonds on the edge.(b) Correlation operators separated by distance $\geq3$ cannot overlap.(c) Two correlation operators overlapping at the lattice sites marked yellow.The weight of the product is the number of red sites in this diagram. The reds represent the non identity operators and the yellow dots show where the correlation operators overlap such that $Z$'s of same index square to identity.(d) In this $5\times5$ lattice, note the minimum weight product of five correlation operators lies on one of the diagonals. }
    \label{fig:my_label}
\end{figure*}
The three different typical 2- products are,

1) 2-products where both the correlation operators belong to same layer
\begin{eqnarray}
\begin{array}{ccccccccccccccc}
 & X & Z & Z & . & . & . & Z & Z & I & I & . & . & . & I\\
\times & Z & X & Z & . & . & . & Z & I & Z & I & . & . & . & I\\
= & Y & Y & I & . & . & . & I & Z & Z & I & . & . & . & I
\end{array}
\end{eqnarray}
 $\implies$ weight,w $=4$ 

2) 2-products where the correlation operators are connected but belong to different layers
\begin{eqnarray}
\begin{array}{ccccccccccccccc}
 & X & Z & Z & . & . & . & Z & Z & I & I & . & . & . & I\\
\times & Z & I & I & . & . & . & I & X & Z & Z & . & . & . & Z\\
= & Y & Z & Z & . & . & . & Z & Y & Z & Z & . & . & . & Z
\end{array}
\end{eqnarray}
$\implies$ $2n$ weight. 

3) 2-products where the correlation operators belong to different layers of the bilayer graph and belong to vertices that are not connected 
\begin{eqnarray}
\begin{array}{ccccccccccccccccc}
 & X & Z & Z & Z & . & . & . & Z & Z & I & I & I & . & . & . & I\\
\times & I & Z & I & I & . & . & . & I & Z & X & Z & Z & . & . & . & Z\\
= & X & I & Z & Z & . & . & . & Z & I & X & Z & Z & . & . & . & Z
\end{array}
\end{eqnarray}
$\implies$ $2n-2$ weight.

The three different typical 3- products are,

1) All three correlation operators belong to same layer
\begin{eqnarray}
\begin{array}{ccccccccccccccccccc}
 & X & Z & Z & Z & Z & . & . & . & Z & Z & I & I & I & . & . & . & I\\
\times & Z & X & Z & Z & Z & . & . & . & Z & I & Z & I & I & . & . & . & I\\
\times & Z & Z & X & Z & Z & . & . & . & Z & I & I & Z & I & . & . & . & I\\
= & X & X & X & Z & Z & . & . & . & Z & Z & Z & Z & I & . & . & . & I
\end{array}
\end{eqnarray}
$\implies$ $n + 3$ weight.

2) Two correlation operators in one layer, third correlation operator in different layer and the corresponding vertex is connected to one of the previous two vertices 
\begin{eqnarray}
\begin{array}{ccccccccccccccccc}
 & X & Z & Z & Z & . & . & . & Z & Z & I & I & I & . & . & . & I\\
\times & Z & X & Z & Z & . & . & . & Z & I & Z & I & I & . & . & . & I\\
\times & Z & I & I & I & . & . & . & I & X & Z & Z & Z & . & . & . & Z\\
= & X & Y & I & I & . & . & . & I & Y & I & Z & Z & . & . & . & Z
\end{array}
\end{eqnarray}
$\implies$ $n+1$ weight.

3) Two correlation operators in one layer, third correlation operator in different layer and the corresponding vertex is NOT connected to one of the previous two correlation operators 
\begin{eqnarray}
\begin{array}{ccccccccccccccccc}
 & X & Z & Z & Z & . & . & . & Z & Z & I & I & I & . & . & . & I\\
\times & Z & X & Z & Z & . & . & . & Z & I & Z & I & I & . & . & . & I\\
\times & I & I & Z & I & . & . & . & I & Z & Z & X & Z & . & . & . & Z\\
= & Y & Y & Z & I & . & . & . & I & I & I & X & Z & . & . & . & Z
\end{array}
\end{eqnarray}
$\implies$ $n+1$ weight.

The minimum weight term from the set of 1-products, 2-products and 3-products is equal to the minimum of $[(n+1),4,2n,2n-2,(n+3),(n+1),(n+1)]$ which is equal to minimum of $[4,(2n-2),(n+1)]$. For $n\geq3$ this minimum is 4. That is any 2n qubit system with $n\geq3$ that follows this bilayer construction is 3-uniform.

\subsection{\label{sec:level3}4 - uniform states:}

Consider a 2D lattice graph state as shown in fig \ref{fig:my_label} with periodic boundary conditions imposed along both the independent directions hence forming a torus like geometry. We claim that,\\
\\
\textit{2D lattice graphs with lattice dimensions more than $5\times5$ and periodic boundary conditions are 4-uniform.}\\
\\
All correlation operators (which are nothing but 1-products) has weight = 5 for such a graph, as each lattice site in the graph has four neighbours. Therefore the maximum entanglement that we can expect from this arrangement is $4\ln2$. By theorem 1 we need to find out the minimum weights of 2-products, 3-products and 4-products to confirm this $4\ln2$ entanglement. If we consider k correlation operators that are far apart(they have no overlapping vertices in the graph), the weight of such a k-product will be $5k$. Like in figure \ref{fig:my_label} (b)with two correlation operators that are three distance \footnote{The distance between two vertices on a graph is equal to the number of edges on any of the shortest paths between the two vertices} apart.

What can reduce the weight of a k-product from $5k$ is vertices common between multiple correlation operators. 

These common vertices correspond to $Z$ operators with the same index squaring to Identity($I$). For correlation operators to have common vertices, the distance between them along this lattice graph needs to be at most 2. So when attempting to evaluate the minimum k-product we need to consider all configurations of k operators such that they are all within two distance of each other. Weight of 2-product = $2\times5-\theta\times2$,
$\theta=$number of overlapping vertices. In the 2D lattice it is possible to exhaustively show that the best way to arrange a 2-product to get minimum weight is to place them diagonal to each other(figure \ref{fig:my_label}(c)).

This produces two overlaps and so the weight is $10-4=6$. This extends for 3-products, 4-products and so on. That is, the minimum 3-product consists of the product of three correlation operators along the same diagonal, the minimum 4-product consists of the product of four correlation operators along the same diagonal and so on. The minimum k-product will involve k correlation operators along a diagonal. If k is less than both lattice dimensions then there will be $2(k-1)$ overlaps or a reduction of $4(k-1)$ weight. Therefore the minimum k-product has weight = $5k-4(k-1)=k+4$. Since $k\geq1$, the minimum weight of 1-product$=5$ $\leq$2-product$\leq$3-product and so on and so forth. Now we need to evaluate what minimum size of the lattice will allow 4-uniform states. Consider a k-product organized along a diagonal where $k=min(l,m)$ where l,m are the number of vertices along the two independent directions, this gives us two more overlaps($-4$ weight) and therefore the weight is $k+4-4=k$ like in figure \ref{fig:my_label}(d).

We need this k product to have weight $\geq5$. This argument can be applied to both dimensions($l,m$) of this lattice and since this lattice has $l\times m$ qubits with $l,m\geq5$, the smallest lattice is a square with $l=m=5$. By this construction we need at least 25 qubits to construct a 4-uniform state.

\section{\label{sec:level3}Conclusion:}
We demonstrated graph states of many different total number of qubits that are 1, 2, 3 and 4-uniform states. Since the structure of a graph state can tell us what quantum circuit can be used to construct the state starting from a product state \footnote{ The qubit at each vertex is initially prepared in state $\frac{1}{\sqrt{2}}(|0\rangle+|1\rangle)$. An edge between vertex i and j corresponds to a controlled-Z gate acting between qubits i and j\cite{hein2006entanglement}}, hence our findings provide a series of explicit recipes to construct these states starting from product states. 
\begin{acknowledgments}
S.S. acknowledges IISER Kolkata, India, for support in the form of a fellowship. S.D. would  like to acknowledge the MATRICS grant (Grant No. MTR/ 2019/001 043) from the Science and Engineering Research Board (SERB) for funding. 
\end{acknowledgments}

\bibliography{references.bib}

\end{document}